# Unblind the charts: Towards Making Interactive Charts Accessible in Android Applications

1st Ishan Joshi
*Faculty of Information Technology*
*Monash University*
Melbourne, Australia
ijos0001@student.monash.edu

***Abstract*—Smartphones are a crucial aspect of routine life in the modern world, and viewing information graphics such as charts becomes common practice for many unassuming tasks. However, for the vision impaired, accessing graphical material presents many difficulties. Android smartphones usually come preinstalled with Google Talkback as a default screen-reader, which attempts to cater for the visually impaired by providing supplementary auditory information when interfacing with supported applications. Still, the crux of this situation is that screen-readers rely on developers correctly implementing the required accessibility guidelines for UI elements, such as charts. Unfortunately, according to the empirical study, more than 88% of the charts found in real-world Android applications are inaccessible to a vision-impaired user, contributing to the wider accessibility issues faced by vision impaired users of smartphones. These accessibility issues can be attributed to a knowledge gap in considering possible disabilities for users, and time costs for developers. To overcome these challenges, this study proposes CAM (Chart Accessibility Module), which aims to reduce time and bridge the knowledge gap required to implement chart accessibility. CAM has two steps, generating chart summary using raw data and feeding it to the screen-reader using the Android Accessibility API for MPAndroidChart library. The user study results show that CAM significantly reduces difficulty and time taken to implement accessibility for application developers.**

## I. Introduction

Smartphone native applications are the most common method of accessing information from the internet [1]. Widespread ownership of smartphones bridges means that they connect us to the world, not only in terms of convenient access to infinite information, but we also heavily rely on them to manage various crucial aspects of our lives.While there are many different forms of information that we ingest through these devices, graphical material such as charts are an everyday occurrence. Native applications [1] which show weather forecasts, stock trackers and health monitors improve the lifestyles of many people. These applications often show changing data through the use of interactive charts. However, smartphones have a naturally heavy reliance on sight. According to the World Health Organisation (WHO) approximately 2.2 billion people live with vision-impairment, of which 36 million are considered legally blind [2]. Unfortunately this presents an incredible challenge for the vision impaired.

To cater for vision impaired users [2], Google Talkback [3] comes preinstalled on most Android devices. The role of the screen-reader is to interface between applications and users, providing spoken feedback that allows a user to build a mental map of what is being displayed to them [4]. The screen-reader is an important tool as it represents the "eyes" [5] that bridges the gap between content and users. Despite their usefulness, the effectiveness of this aid relies on content description for UI elements to be included in development by application developers [6] as a prerequisite.

An inaccessible chart is undetected by the screen-reader when browsing the UI elements. Inaccessible charts can cause difficulties for users without vision. Ensuring that everyone has access to information is not just a moral obligation, but in certain cases required by law [7]. Overcoming these accessibility issues will result in many benefits such as social inclusiveness, empowerment and social justice [8] for the vision impaired. Typically users utilise a scribe or look for volunteers to assist with understanding charts, however such methods are time consuming and result in information delay. Furthermore, it limits the extent to which users can form their own interpretations, subjecting them to information bias [9]. Following HCI principles, accessible charts take a step towards user-sensitive and universally inclusive software design [10].

Currently, vision-impaired users rely on a screen-reader to understand charts presented in applications. Interactive charts [3] are commonly found in health & fitness, finance, and weather applications. These fundamental applications enrich users' lives and form an essential component of well-being. However, 88.1% of the charts tested during the empirical study are inaccessible using a screen-reader. Additionally, the vision-impaired users may be more eager to try new applications in search of accessibility support [5], which provides additional motivation for application developers by implementing accessible charts to distinguish themselves from competition.

Charting libraries are a well-tested framework for developers willing to implement charts, which saves time and effort. A study of 11 popular Android charting library sample applications concluded that all are not accessible. Charts are non-standard UI elements, often termed as custom-views. A custom-view usually requires implementation of "UI cues" [11] and commitment from developers to enable accessibility. This is often omitted due to lack of time, regulations, awareness, cost factors and no explicit requirements for success criteria [12].

[1]Applications or apps refer to native Android Applications in all following sections

[2]*Users* implies vision-impaired users where these are used interchangeably in subsequent sections.

[3]*Chart* refers to interactive charts

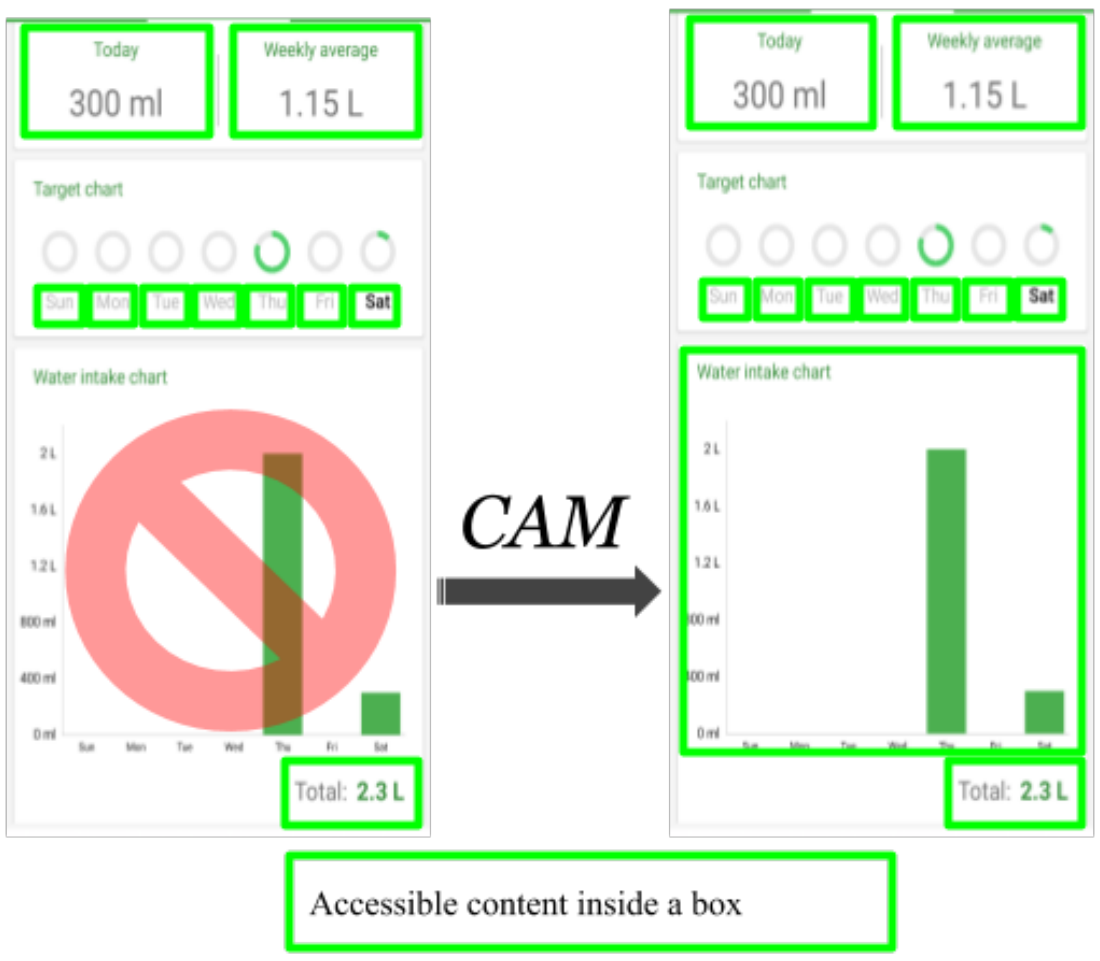

Fig. 1: A visual representation of screen-reader traversing UI elements on a screen to generate description. Green-boxed elements are focusable and can generate a description. CAM implements accessibility guidelines for custom-view to make charts accessible.

In the real-world applications, lack of accessibility in charts is evident through the findings of the in-depth empirical study conducted as part of the study. Out of the 151 charts that are manually tested, the results show that only 11.9% (18) charts are partially accessible, the other 88.1% (133) are inaccessible due to no UI cues being implemented, confirming the poor state of accessibility of charts in Android applications.

This study proposes CAM (Chart Accessibility Module), a novel approach focused on using raw data and chart context to generate descriptions for line, bar and pie charts. It helps developers to create accessible charts by reducing the time and knowledge barriers developers face with accessibility implementation.

A controlled user study of two groups found that CAM developers are always successful in implementing an accessible chart versus a 55% completion rate without it. Developers using CAM also experienced 58% improved time efficiency and rated the difficulty 2.1 lower on a five-point Likert scale. The study demonstrates that CAM is highly effective in bridging the knowledge gap and reduces the time taken to implement accessibility to the currently available options using MPAndroidChart (most popular) library as the subject.

The contributions of this study are summarized as:

- To the best of knowledge, this is the first study undertaken which measures charts accessibility in Android applications.
- We propose CAM and added accessibility features to MPAndroidChart library which reduces development time when implementing accessible charts.
- Implemented an approach into the practical tool for assisting developers in the real world.

Section 2 discusses the background around chart accessibility with a focus on MPAndroidChart and the role of screen-reader. Past-works in the domain of static chart accessibility are also described. Section 3 presents an empirical study on charting library distribution and accessibility of charts in apps. Section 4 introduces CAM module and the implementation. Section 5 shows results of user study evaluation. Section 6 provides the conclusion and future improvements.

## II. Background

### A. Charts Implementation

The Android Software Developer Kit (SDK) does not provide any built-in charting components. As a result, applications developers adding a chart usually rely on third-party libraries, many of which are open source.

All Android UI components extend from the base View or ViewGroup class. View provides a canvas-like abstraction [13]. A canvas allows each pixel to be independently rendered using components like Paint and Brush. Android Developer Documentation suggests custom views are an ideal candidate for drawing charts [14].

| Library | Stars | Base Class | Accessible? |
|---|---|---|---|
| *MPAndroidChart* | 31,800 | ViewGroup | No |
| *AChartEngine* | 708 | View | No |
| *AFreeChart* | 7 | View | No |
| *Android Charts* | 1,200 | View | No |
| *Android Plot* | 376 | View | No |
| *Droid Chart* | NA | NA | No |
| *Eaze Graph* | 1,500 | ViewGroup | No |
| *Holograph Library* | NA | View | No |
| *HzGrapher* | 133 | View | No |
| *SciChart* | NA | NA | No |
| *William Chart* | 4,500 | ViewGroup | No |

TABLE I: Android Charting Libraries Data

There are several popular libraries. A list of 11 charting libraries[4] along with their Github start count (if applicable) and base view component is compiled in Table I. Each library differs in terms of the API provided, features available and ability to handle differing types of data. However, all libraries build "on the fundamental" [15] *View* or *ViewGroup* class which allows each chart to be treated as a custom-view in terms of accessibility. Unfortunately custom views are inaccessible due to not being screen reader focusable by default [16].

Initial source code analysis showed all libraries are missing the implementation of accessibility interfaces. Each charting library identified has an associated sample demo application. As part of the study, each of the samples are tested using a screen-reader for accessibility evaluation. All 11 demo applications are inaccessible using a screen-reader. Given that real-world applications use charting libraries as building blocks, charts implemented in other applications may be inaccessible to screen-readers, unless developers specifically override the default behaviour.

[4]Wrappers to Javascript Libraries *Highchart, AnyChart* are ignored

### *B. MPAndroidChart Exploration*

Based on empirical study charting library distribution, AppBrain Charting Library Data and Github stars MPAndroidChart [17] is the most widely used charting library when making Android applications. This section explores the features, common interactions and limitations of the library in the context of accessibility.

Integrating MPAndroidChart is done through two dependency statements in the project's Gradle build file. This process automatically fetches and packages the library into the project. Several documented samples, implementation guides and full JavaDoc [17] are available for all developers.

Line, bar, stacked-bar, column, pie, spider, candlestick, scatter and bubble charts are well supported. The extensive range allows for flexibility in terms of choices during implementation. Each variant supports full styling configuration. Colours, point shape, size, animations, etc… can be controlled by the developer to control the presentation of the charts. Gestures like click, long-click, pinch and rotation are handled encourages chart exploration for sighted users. Additionally, the charts support adding chart title, axis labels and legend to contextualize the chart.

MPAndroidChart library charts are abstracted into two distinct layers; presentation and data. The presentation layer stores all view configuration, which is used to redraw the chart by painting each pixel on a virtual canvas. Presentation layer's view configuration, like line color and width is not relevant for a user relying on a screen-reader to read out text.

Data layer is abstracted by providing datasets for each chart. Each dataset stores a list of entries corresponding to each data entry in the chart. This abstraction ensures that access to raw data is always contained within the chart instance drawn. As the data can change over time, the corresponding dataset list can be modified by the client code. Once the data is updated, MPAndroidChart must be notified using *notifyDatasetChanged()* and *invalidate()* Android View methods, which trigger the redraw of the chart. This separation allows data-based views to update dynamically due to the underlying data and view configuration. Accessibility description generation only requires access to the underlying data where the presentation layer configuration can be omitted.

### *C. Screen Reader and Chart Accessibility*

Talkback is a preinstalled screen-reader on most Android devices as part of Google's accessibility suite. It allows vision-impaired users to access and interact with content displayed on the screen. Users can navigate using swipe gestures to find content on the screen. Navigation modes like control, links, heading and default are supported to help the user navigate content quickly. Screen-readers work by traversing and parsing the UI view hierarchy to focus on appropriate elements and read a description.

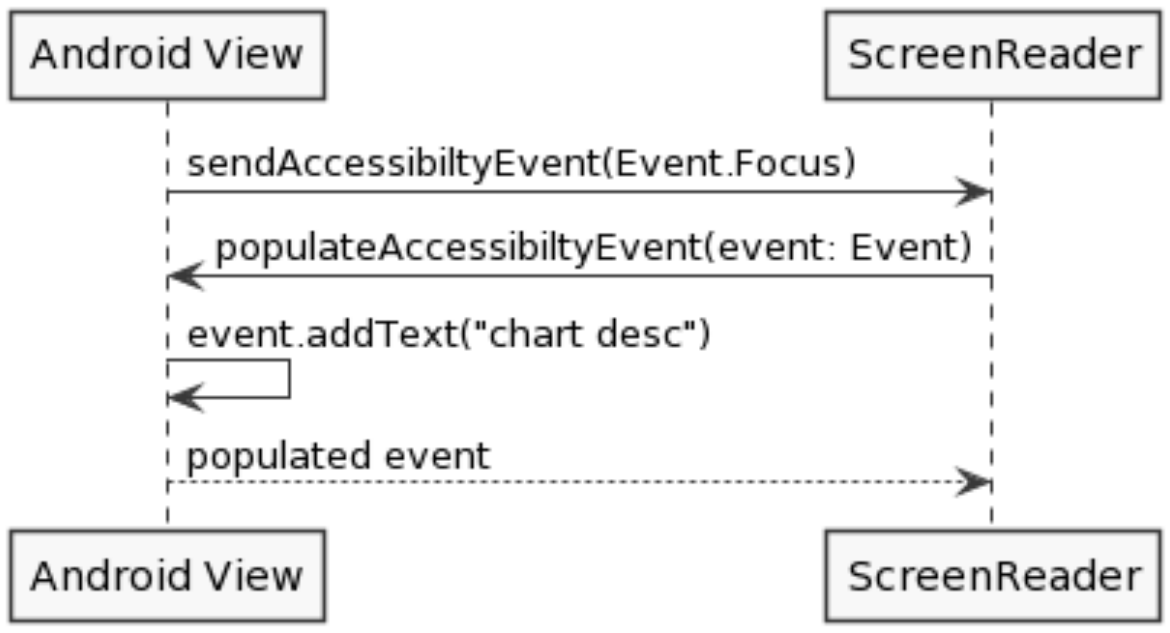

Fig. 2: Sequence of events that make views accessible

Each UI view subclasses the Android View class. Three important View attributes determine the accessibility support with screen-readers; *focusable*, *screen-reader-important* and *important-for-accessibility*. *Focusable* is responsible for allowing the screen-reader to detect the view as actionable. These attributes must be set to appropriate values for the screen-reader to discover and focus on the custom-view chart [16]. As shown in Figure 2, each UI view is also responsible for *sending* [14] and *populating* [15] accessibility events to provide constant feedback to the screen-reader. Events can be populated with text which is read out by the screen-reader service.

Many primitive SDK elements like *Button and TextView* implement accessibility behaviour by default. However, for custom-views, the focus and click events are automatically sent if the correct accessibility properties are set. Focus event is an ideal candidate for populating the accessibility event chart overview description. Each custom-view chart has access to the raw-data, which can be used to generate a textual description for the screen-reader.

### *D. Accessibility Testing*

Accessibility Scanner [18], UIAutomator Viewer [19], static linting and Firebase Test-Lab [20] are available to developers to test for accessibility performance of an application. Automated tools rely on focusability of UI elements, which is not implemented by default and is required as a prerequisite for automated accessibility testing. It relies on the developer setting the correct values for accessibility view attributes by overriding defaults. Subsequently, the responsibility of testing the accessibility of custom-view charts falls on the developer. Manual testing using a screen-reader is recommended in developer guidelines [21] to find missing descriptions and inconsistent navigation for custom-views.

### *E. Related Past Works*

Studies in the past focus on describing data tables, static charts and alternate data presentation through sonification.

All types of chart data can be semantically represented as a table. Tabular data allows charts to be described using key values to construct meaningful descriptions [22], [23]. Template filling [24] is also described as an alternative to creating summary textual description. Data table descriptors rely on raw data to generate description.

Static chart descriptors have been commonly studied to help with alternate text generation for chart images. An ideal use case scenario is an image based chart shown on websites, where automated descriptors can add alternative text for untagged charts. In order to create descriptions, these systems rely on chart type detection and data extraction through machine learning models. Chart-Text [25] achieved 78.9% accuracy for chart data extraction from images. A recent study surpassed "the capabilities of the average human annotator" [26] for non-changing charts.

Additionally, sonification of charts overcomes the limitation of presenting just text output as the ear has a "high power of discrimination" [27], just like sight. Guidelines for sonification of charts are published where MIDI tones can be mapped to y-axis values of data [28]. Although sonification results in a better user experience for certain types of charts, the study focuses on chart integration with screen-readers, where sonification is seen as the next obvious improvement.

This study builds on two key areas of improvement; access to raw data and describing dynamic charts where the underlying data can change. Access to raw data improves the accuracy and quality of descriptions. Being able to describe changing charts is highly applicable for use with charting libraries, an improvement over static images.

## III. Empirical Study of Charting Accessibility in Real-World Android Apps

This empirical research explores the distribution of charting libraries and the general accessibility of charts presented in real-world apps. It provides statistics backed by exploration of real-world applications, which highlights the poor-state of chart accessibility as the motivation for developing CAM. Two key research questions can be formed to gain insight into this topic.

***RQ1**: What is the most common charting library used in real-world Android applications?*

***RQ2**: To what extent are the charts found in real-world applications accessible?*

### A. Charting Library Usage in Android Apps

To analyse charting library usage in real-world applications, the following procedure is used for data collection and analysis; *Identify categories likely to contain charts → Search and download APK files → Analyse APK for presence of charting library → Findings.*

The categories narrow down the search providing targeted focus. Using AppBrain data to source keywords of application names which are known to have charting libraries narrow the search. APK files are downloaded to enable static analysis.

*1) Categories Identification:* Google Play store is a popular app-store, which distributes Android applications. The store is divided into 35 application categories. Based on data from AppBrain Charting Library Analysis [29] and intuition, Finance, Weather, Health & Fitness and Travel are identified as categories in which applications are likely to contain charts. Intuition rules filter out categories like Comics as they are unlikely to contain charts. AppBrain charting data only considers the presence of only three charting libraries, which is a limitation of the dataset. A list of categories provides targeted focus for finding applications with known charting libraries.

*2) Sourcing APK Files:* Android APK is a common archive format that allows packaging and distribution of Android applications. Each APK contains a manifest file, resources folder and compiled DEX classes generated from source code, which includes third party libraries used. APKs can be decompiled for static analysis and can also be easily installed onto a compatible device.

Python Google Play Unofficial API [30] is used for two key steps; targeted search queries and downloading APK files. The keywords are based on AppBrain data as well as top applications in a given category likely to contain a chart. For each keyword, a depth-first search based approach is undertaken to yield similar applications. For example, the keyword 'stock' yields five other terms; 'stock market', 'stocks for beginners free', 'stocks', 'stockx' and 'stocks app'. The APK files for search results are downloaded, while preventing duplication. Additional APK files for the likely categories are sourced from ANZHI (Chinese) App Store with help of other researchers. A total of 1911 apps and respective APK files are downloaded for further analysis.

*3) Library Detection:* All user and included library code is compiled into DEX classes which is packaged with the APK files. However, finding a specific library within an application is challenging due to Android Obfuscation [31], which hides all code and method signatures. LibRadar [32] is an obfuscation resilient utility tool that can perform APK analysis to extract a list of known third party libraries. It takes an APK file as input and produces a list of third-party libraries found as output.

LibRadar relies on a dataset of package names of libraries to facilitate detection. For charting library detection, a large dataset [33] that lists package names and meta-information of popular charting libraries is filtered using the 'chart' keyword. Each known missing library is manually augmented. This results in a complete dataset [5] which can be used with LibRadar to find known charting libraries.

*4) Findings:* After running LibRadar detection on 1911 APK files, 842 applications (44.1%) contained a known charting library. Out of those, 36 apps had more than one charting library detected. MPAndroidChart appears in 65.3% of the applications, followed by AChartEngine and Chartboost.

[5]https://github.com/ish-joshi/cam-chart-accessibility

Answering *RQ1*, APK analysis concludes that MPAndroidChart is the most popular charting library in real-world applications, findings consistent with AppBrain data and Github stars. This justifies the pick of MPAndroidChart as the chosen library for improvements.

A further analysis of the applications with charting libraries is conducted to answer RQ2 which determines the extent of accessibility in charts.

### *B. Accessibility of Charts in Android Apps*

After identifying the presence of a charting library, the next step is to determine if the charts are accessible in the context they appear in. This process involves two key steps; locating a chart and determining chart accessibility.

Finding and locating a specific element requires exploration of the application. There are two main ways exploration can occur; systematic and random.

AppCrawler [34] is an ideal candidate for systematic exploration. It clicks into screens and intelligently attempts to explore different screens within the application. As output, it produces screenshots and an accessibility report. Unfortunately, the report does not include custom-view charts.

On the other hand, ADB Monkey [35] provides unstructured exploration by simulating random events. It is an alternate way to explore apps in a scalable manner. However, it is impossible to replicate the series of events used for exploration again.

All available exploration tools are limited by login barriers, captcha verification and inability to generically perform valid data input into any application. A combination of the listed factors are a hurdle in automated exploration of apps.

A manual study approach is best suited for finding charts and testing accessibility, as it allows the tester an attempt to bypass the barriers listed above. Out of the 842 applications with a known charting library, only the 239 apps downloaded from Google Play Store are manually tested. Applications downloaded from the ANZHI app-store are omitted due to language barriers in the apps.

The manual approach captures a snapshot of many screens in real-world applications. Each snapshot captures a screenshot, XML screen-hierarchy and boolean value for the presence of a chart. If a chart is available, the type of chart and accessibility status is also captured. Data capture and labelling is done using a custom script using Python ADB Wrapper [36], which is available for further collection of data.

A chart is considered accessible if the screen-reader focuses on the chart view boundary or if a relevant description is present in close proximity of the chart-view in an accessible UI element.

90 applications could not be tested due to installation errors. 382 snapshots are captured from 149 applications, which are a mix of charts and non-charts. Several of the 149 applications could not be tested in depth due to barriers such as geo-restrictions, unsupported devices for the APK file, strict login barrier (e.g. banking application) and language barrier.

In total, 151 chart snapshots are captured. Only 11.9% (18) charts are partially accessible using the screen-reader. 88.1% (133) of the charts remain completely inaccessible, as they are not detected when using a screen-reader. This shows a very poor state of charts accessibility in real-world applications.

From the 18 charts that are accessible, through observation these can be categorized into four categories; text within chart, data table, chart-text description and focusable chart.

1) ***Text Within Chart*** (Figure 3) have chart-text placed within the chart itself, using a TextView component, which is accessible. When using the screen-reader, the text-views come into focus and read the text. Custom view that contains the chart is not focusable.

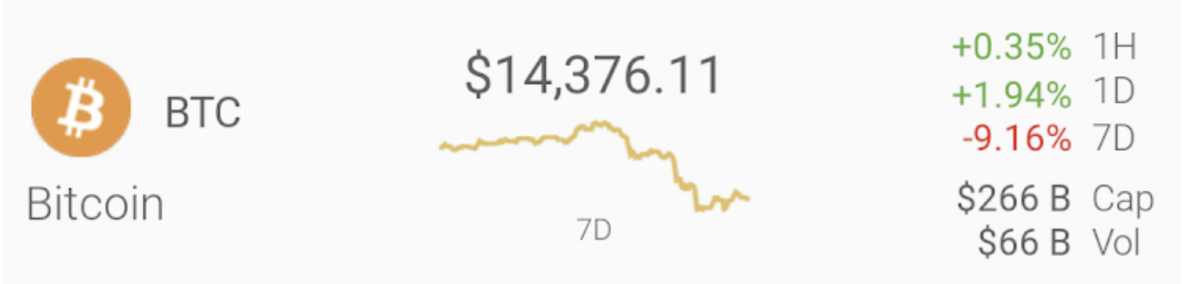

Fig. 3: Accessible text "$14,376.11" shown within the chart

2) ***Data Table*** (Figure 4) charts have a table of values present immediately below the chart. Although the chart custom-view is not accessible, the table of values provides a systematic exploration of each of the data points.

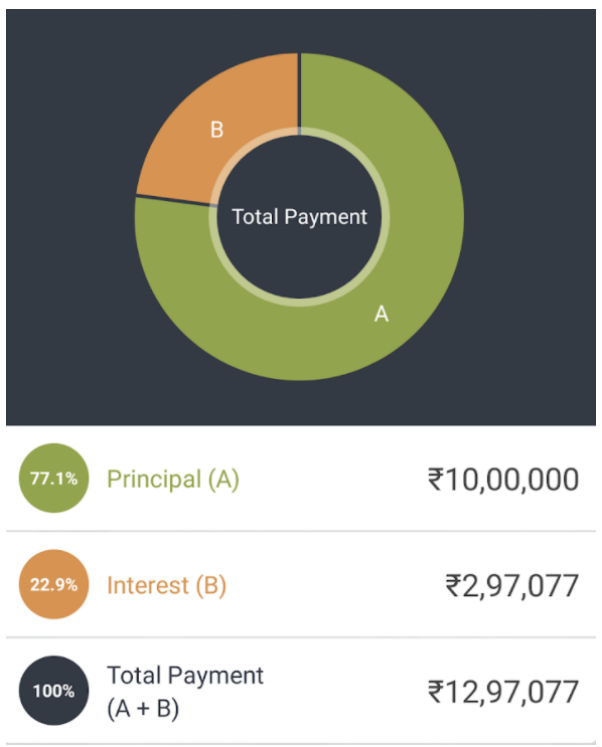

Fig. 4: Data Table allows exploration of list items using a screen-reader

3) ***Chart Text Description*** (Figure 5) provide a textual description about the chart information is shown in close proximity, within a few swipes when navigating using a screen-reader. The custom view containing the chart is inaccessible, but the text below it is read out by the screen-reader.

4) ***Focusable Chart*** allow screen-reader to detect the custom view which draws the chart. This can be achieved if one of the accessibility properties are set correctly for the custom-view.

*TextView* is a vital UI component for accessibility of observed charts, as the content within them is accessible by default. All observed charts do not provide any insight into underlying data, where the meaning is lost for vision-impaired users. Having certain accessibility text is a step towards accessible charts, but complete summary description is expected to make the charts acceptable.

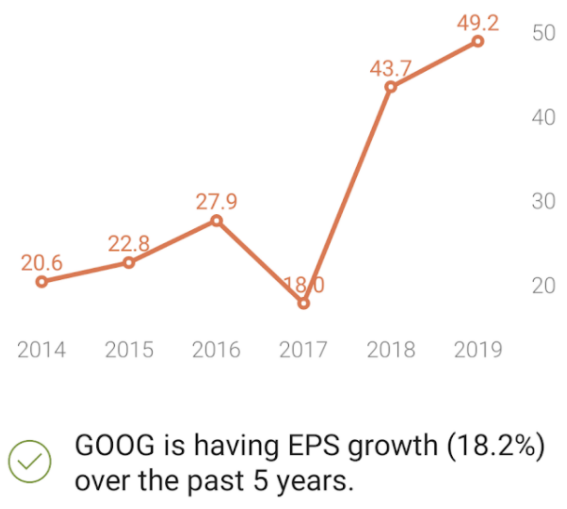

Fig. 5: Accessible chart summary text describing chart features is provided below the chart

Answering *RQ2*, 88.1% of the observed charts are inaccessible. Moreover, the accessible charts are only partially accessible, as no meaningful information about the underlying data is communicated. The state of accessibility in charts is *poor* considering the majority of charts are completely ignored by the screen-reader.

### C. Automated Chart Testing

After completing the manual data capture, certain heuristics are developed through observations of XML view-hierarchy in screens containing charts.

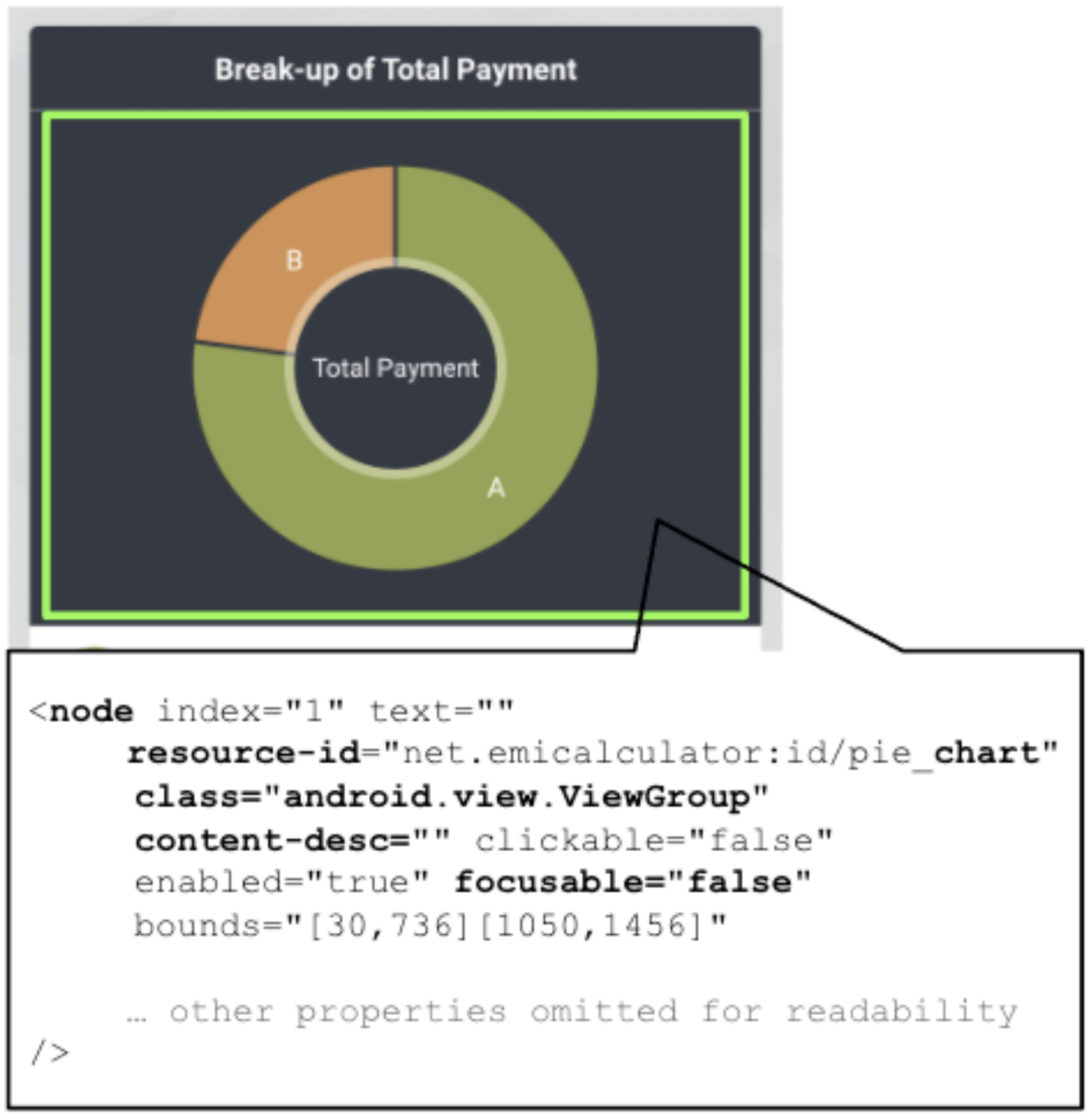

Fig. 6: XML Node for a chart-view showing relevant accessibility attributes

XML UI View-Hierarchy (*UI Dump*) Figure 6 is a capture of the elements visible on the screen. It captures the *bounds* of the view, the *focusable* attribute, *content-description*, *resource-id* and the *class name* of the view alongside other meta-information. Captured XML masks the class-name of *com.github.mikephil.charting.charts.PieChart* as *android.view.ViewGroup*, which cannot be used in isolation to accurately find a chart. However, from observation, if a node class is *View, ViewGroup, SurfaceView or FrameLayout* and the resource-id contains the phrase *'chart'* it is likely a custom-view with a chart. A node is considered accessible if it is *focusable* or has a *non-empty content-description*.

Automated analysis of XML view-hierarchy data achieved 81.6% accuracy in correctly outputting presence or absence of a chart, while achieving 86.1% precision and 75.6% recall. This heuristic can be used in further studies to locate charts and test for accessibility, for a larger scale exploration in the domain of charting accessibility.

### D. Data Summary

1911 apps APK files are downloaded for analysis. APK file analysis for a third party library using LibRadar found 842 apps packaged with a known charting library. MPAndroidChart is the most commonly occurring charting library. 36 apps also had multiple charting libraries. Finding charts and testing accessibility cannot be automated due to tool limitations. Manual testing is conducted on 239 apps downloaded from Google Play Store which have a charting library.

From the manual study of 239 applications, 90 could not be tested due installation errors. 149 apps produced 382 snapshots which include XML dump, screenshot and meta information capturing the presence, type and accessibility of the chart. 38 bar, 88 line and 25 pie charts total the 151 charts found. 88.1% (131) of charts are skipped by the screen-reader and 11.9% (18) of the charts are restrictively accessible.

Through heuristics, an automated XML analysis approach which can scan for chart presence and accessibility on a given is devised. This approach reached 81.6% accuracy and can be potentially used to automatically identify charts for future studies.

These findings from the empirical study confirm the lack of accessibility for charts found in real-world applications, which motivates the study to develop an approach that makes creating accessible charts easier.

## IV. Approach

This study proposes CAM (Chart Accessibility Module) to help developers implement accessible charts to overcome the poor accessibility of charts observed in the empirical study. It aims to reduce developer effort and time taken. Two key aspects of the approach are generic implementation and future extensibility. MPAndroidChart is used for proof of concept due to its popularity and open-sourced availability. However, the approach is applicable to any charting library.

Having access to the raw-data and meta-information improves content description quality. CAM provides an easy mechanism to feed textual summaries to the screen-reader, as well as a tool to generate descriptions based on the domain of the chart; e.g. weekly rainfall forecast, stock trends for a day and pie charts.

MPAndroidChart has several interactions possible in a single view class. In order to enable interactions with smaller components like points or pie slices, a "virtual view hierarchy" [37] is a recommended approach, which allows the user to perform fine grained actions. These can be seen as an obvious extension of this study, however for now, CAM only focuses on providing only a summary of a chart.

This can be seen as an obvious extension of this study, but only summary description is in scope for the current CAM proposal. Two key steps are described in the approach that enable chart accessibility; generating chart summary and screen-reader integration for MPAndroidChart.

### A. *Charts Summary Templates*

Three types of charts are considered for template creation; line, bar, column and pie chart. The summary of any given chart depends on two key factors; type of chart and domain of the chart. Pie charts show proportions, whereas line charts usually show a trend over time. Context setting, data description and analysis make a good chart summary [38].

Inspiration for template creation is taken from past papers, IELTS chart-description coaching and generalization from hearing news reports. Context setting involves reading out the title, range of data and the units if applicable. Data description has two approaches; reading each data-point or a trend summary. Data analysis aims to infer the intended meaning of a chart, providing reasons for the data-values. All three aspects can be generated using the raw data for any underlying chart. However, generating analysis is based on creating meaning from data, which can cause copyright issues even if the information is changed for better content delivery [39]. Therefore, the templates will not make any inferences of data which also prevents bias in information.

*1) Pie Charts:* Title, categories and the number of points are important for pie-chart context setting for a pie-chart [37]. Proportion of each category is sufficient to generate a data-description. To make compression and comparison easier the values are read in descending order [38]. Reading each data point is an efficient way to describe pie-charts, where a template fills in the category title and the relevant proportion [39]. A limit of seven [40] entries is applied as it is known to overwhelm the short-term memory of a user. Unread entries are grouped into an 'others' category. Additionally, repeating the same structures for reading each data point should be replaced with varying structures [38], like replacing 50% with 'half' for better comprehension. Using the rules, a generic pie-chart template is formulated.

*2) Bar Charts:* There are two observed categories; bar and column charts. A key difference is the values stored in the x-axis. A discrete value which does not represent a temporal component is considered a bar chart; e.g. Car model sales counts. A column chart has a temporal component, which shows change in values over time, where the information about the trend is important.

Bar-charts can be treated as a pie chart where the proportion is replaced by raw counts for each category, as a simple substitute. In contrast, time based charts involve parameters like minimum, maximum, average values, rate of change and trend detection. A template is created for weekly-forecast of rainfall as an example.

Describing rainfall grouped per day involves reading out the day and the corresponding forecasted value. Weather phrases like 'light-rain', 'moderate-rain' and 'downpour' can be used to create additional meaning [41]. This is an example of customized templates based on units and the context in which a chart appears. Additional templates can be created for any domain; fitness tracking, daily temperature or wind-forecast.

*3) Line Charts:* Health & Fitness, Weather and Finance apps predominantly show line charts. From the empirical study, stocks commonly show a trendline showing fluctuations in prices over time. A template for describing stock is created, using Yahoo Finance [42] app as inspiration. Start-value, end-value, low, high and trend values along with respective timestamps form a key component when generating chart summary.

### B. *Screen Reader Integration*

Generating a description of a chart is the first step towards integrating with a screen-reader service. Accessibility hooks must be implemented on the custom-view. MPAndroidChart is modified to enable communication between screen-readers and charts via the Accessibility API. It is done by setting the focusable property to allow focus. This approach is not limited to MPAndroidChart, but it can be generalized to any Android, iOS or Javascript library.

```
class Chart {
    Focusable = true // screenreader focus

    populateAccessibiltyEvent(event) {
        // Feed summary to screen-reader
        event.addText(chartDescription())
    }

    abstract String chartDescription()
}

class PieChart extends Chart {
    String chartDescription() {
        // Current Chart data
        String[] xVals; Float[] yVals;
        String title = 'OS␣popularity'

        // CAM Module
        PieDescriptor descriptor =
        PieDescriptor(xVals, yVals, title)

        return descriptor.describe()
    }
}
```

To feed text to the screen-reader, the accessibility event is populated with a description generated using CAM descriptor. A descriptor generates string description using primitive data. The current system supports PieDescriptor, StockDescriptor and RainfallDescriptor, which create specialized output. Each chart implementation is responsible for generating a description based on the dataset using the CAM module, which automatically populates accessibility events with summary text. Loose coupling between CAM and MPAndroidChart enables future extension to any library.

Additionally, contextual information about the units and the data range can help create more meaningful descriptions. A chart describing wind can reveal additional information using the knowledge of the y-axis values. A numeric speed can be converted to a familiar textual description like 'light wind' or 'gust' using a Beaufort Scale [43] lookup. The descriptor framework allows data to be taken to generate a textual description. It promotes future extensibility and gives developers flexibility by providing a plug and play swap system custom descriptors.

```
interface IDescriptor {
    String describe()
}

class PieDescriptor implements IDescriptor {
    public PieDescriptor(x, y, title) {}

    String describe() {
        // use x, y and title to create
        // template summary
    }
}
```

Overall, the CAM module and MPAndroidChart modifications simplify description generation and feeding summary text to a screen-reader for application developers.

### C. Example Chart Summaries

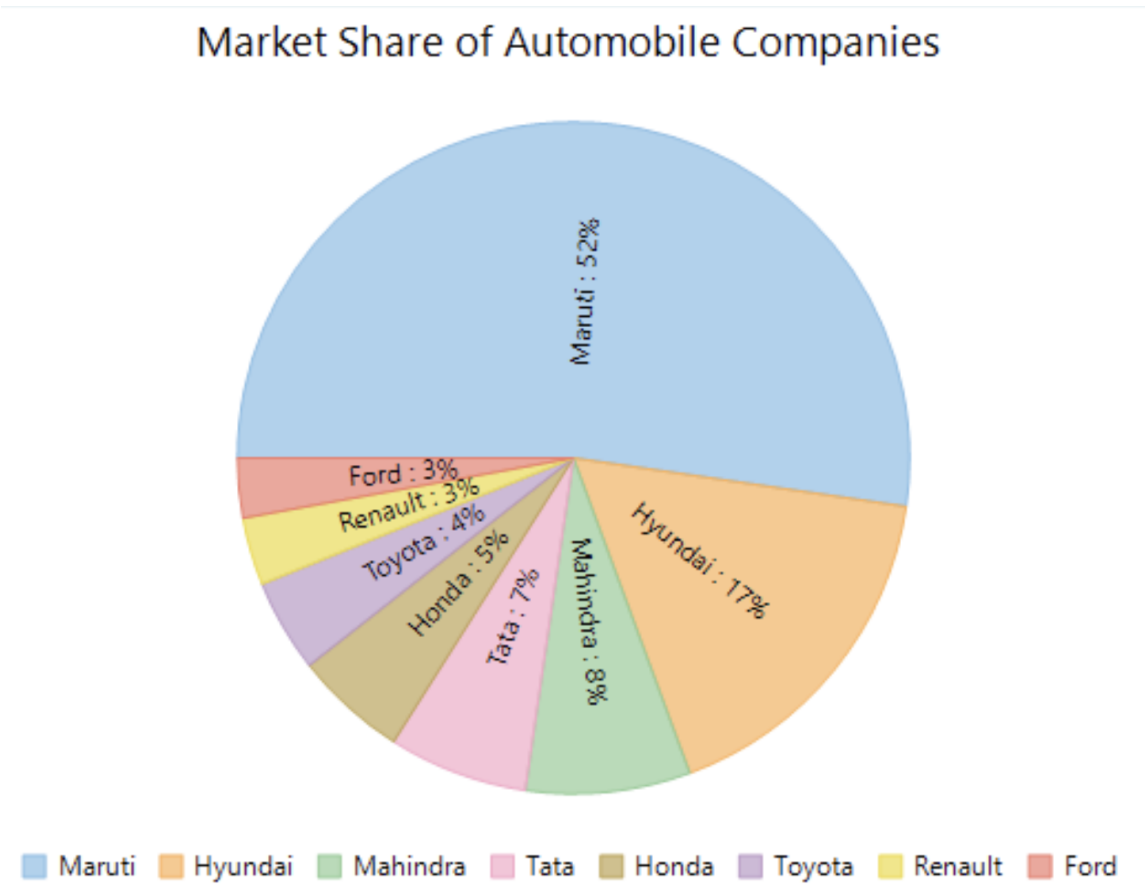

Fig. 7: *Chart Summary:* The pie chart describes Market share of automobile companies. There are 8 data points. Maruti fills up approximately half of Automobile Companies, Hyundai fills up 17.00 percent of Automobile Companies, Mahindra fills up 8.00 percent of Automobile Companies. Tata, Honda, Toyota, Renault and Ford fill up the rest.

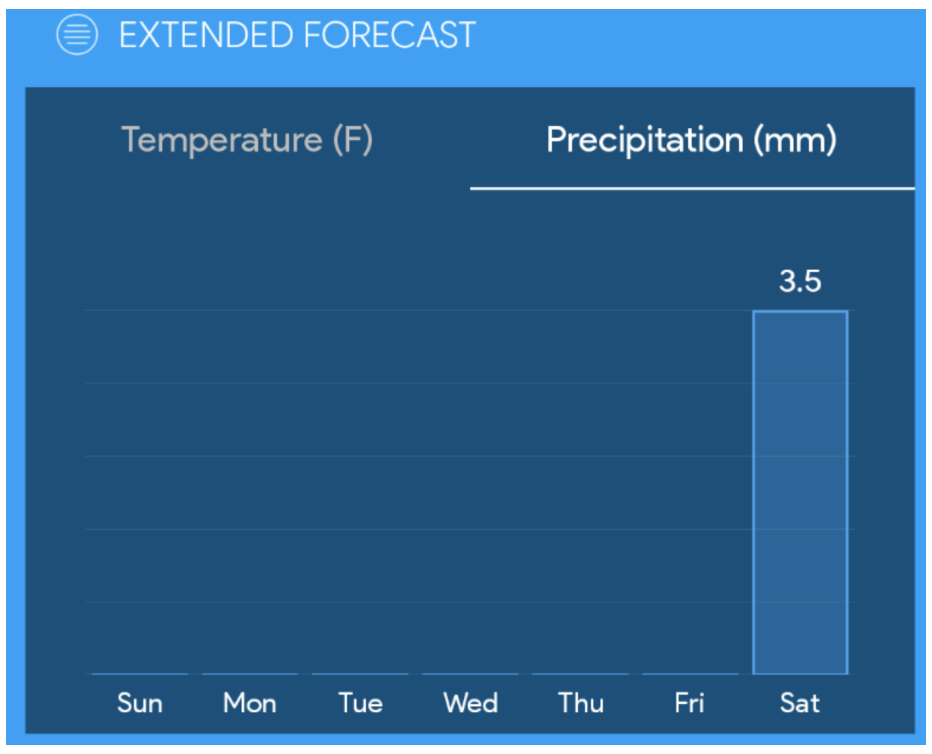

Fig. 8: *Chart Summary:* The bar chart describes the extended forecast for precipitation in Melbourne. The units used are millimeters for rainfall. On Sunday, Monday, Tuesday, Wednesday, Thursday, Friday no rainfall is forecasted. On Saturday, a light-rain of 3.5mm is predicted.

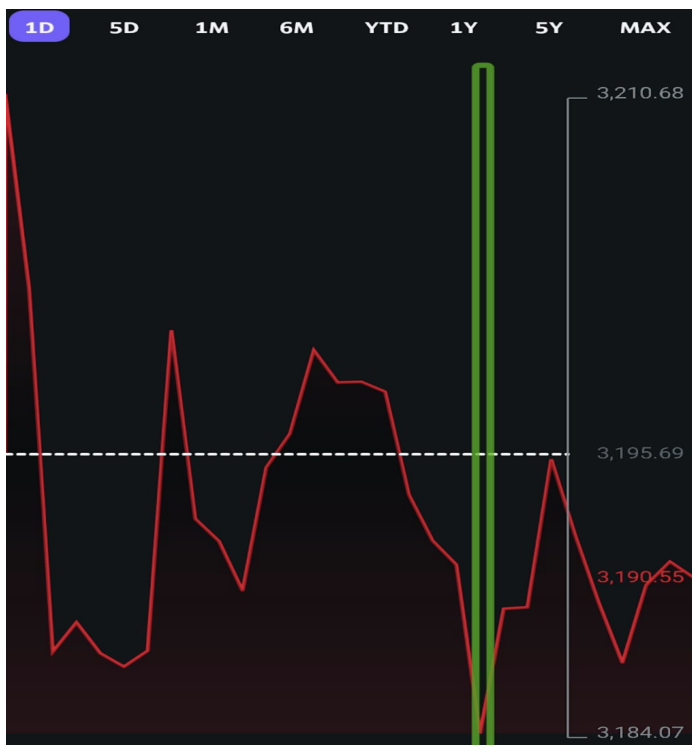

Fig. 9: *Chart Summary:* The line chart shows information about Amazon, which is trending downwards. The chart shows data from 12 Oct 2020 11:41 17 seconds to 12 Oct 2020 15:41 55 seconds. The starting value is 3195.69 US Dollars and the closing value is 3190.55 US Dollars. The minimum value is 3184.07 US Dollars on 12 Oct 2020 15:41 55 seconds. The maximum value is 3210.68 US Dollars on 12 Oct 2020 11:42 16 seconds.

## V. Evaluation

CAM module aims to make accessibility implementation easier, which reduces time taken for an app developer to add accessibility to charts. To confirm the usefulness of our work, a controlled study of two developer groups is conducted. The evaluation focuses on the *time taken*, *success rate of implementation* of accessible chart, *difficulty* and *satisfaction* scores, to show the effectiveness of our approach.

### A. Setup

The study asks developers to implement three accessible charts. Two groups are created to test CAM+MPAndroidChart (test group) versus the publicly available library (control group). Each group is provided with a code shell which implements a pie, bar and a line chart, which shows how to extract the data from MPAndroidChart library chart. Furthermore, all participants are given an introduction to using Google Talkback to test accessibility. The test group is provided with the CAM module and documentation as additional resources. Both groups are tasked to make the charts accessible with specific instructions to refrain from using hard-coded data.

The pie chart plots automotive market share proportions for car brands. The bar chart shows the forecasted rainfall in the upcoming week. Finally, the line chart shows stock values for Amazon.

As part of the study, six developers are recruited. To ensure fairness in terms of skills and experience, four junior (less than two years experience) and two senior developers (more than ten years experience) are split into two even groups. Each group contains two junior developers and a senior developer.

Each developer is only allowed a maximum of 20 minutes for each chart as a base line. Additionally, the participants are asked to attempt the pie chart first, followed by the bar and the line chart, which maintains consistency. As developers gain familiarity with the available tools and API, the subsequent attempts take lesser time. Doing the tasks in a specific order mitigates a source of potential bias. All participants carried out the tasks independently without any discussion with one another.

Once the tasks are completed, each developer provides a rating. Participants provide ratings on a 5 point Likert scale [44], [45] for difficulty and satisfaction score. The difficulty score is captured where 1 is the most difficult and 5 being the easiest. Satisfaction score is captured where 1 is not-satisfied and 5 being totally satisfied. Participants also provide some informal feedback about the experience and CAM module after completing the study.

As there is no ground truth for accessibility description. As a result, an additional developer (denoted as E1) is recruited as an evaluator. E1 has prior experience in accessibility for Android applications. Instead of a discrete value of acceptable or unacceptable, a new metric of acceptability score is introduced for each of the charts. The human evaluator will assign a score where 1 is unacceptable and 5 being acceptable.

### B. Results

The two groups have significant differences in the task completion rate and the performance in terms of time taken. The control group only completed 5/9 tasks amongst the three developers, whereas the test group achieved 100% completion rate.

Significant improvements in terms of the time taken to complete the tasks are observed. The control group took 18.3 minutes on average per task, in comparison to 10.7 minutes spent by the test group to implement accessibility to charts. This takes into consideration all the tasks average, not a specific type of chart. Overall, both groups took 10% longer to complete the pie chart compared to the other two charts. This can be explained by improved familiarity with Android and CAM documentation after exposure to the first task.

With the use of the CAM module and MPAndroidChart modification, developers spent 58.1% less time and achieved a higher completion rate, which shows the effectiveness of the CAM implementation.

There is a gap between senior and junior developers due to experience. In the control group, junior developers spent 2.5 minutes more on average than senior developers. In contrast, for the test group, the performance of junior and senior developers is comparable. This result shows that our approach bridges the gap in knowledge caused due to lack of experience.

Ratings for satisfaction and difficulty level are captured on a 5 point Likert scale. The test group rated their satisfaction of completion 2.3 points higher than the control group. Difficulty level has a difference of 2.1 between the control and test groups. The test group found it less difficult to implement accessibility, compared to the control group.

All results suggest that CAM achieves many benefits for application developers; it makes adding accessibility easier, more satisfying, higher completion rate and time savings. A Mann-Whitney [46] test was performed between the values captured for completion rate, time taken, satisfaction and difficulty ratings. All p-values are less than 0.01. This approach significantly improves upon time taken and bridges the gap between junior and senior developers.

Test group acceptability score is rated as 3.4 by the evaluator (E1), versus 1.3 given to the control group. Some charts received a score of 1 in the control group as the developer is not successful in implementing any accessibility on the charts.

As part of informal feedback, a participant mentioned that they found the pie chart difficult as it was the first chart, in the succession of three. The control group developers who weren't familiar with custom views accessibility found the Android documentation hard to connect with MPAndroidChart, in terms of populating accessibility events. The senior developer in the control group mentioned the idea of altering the library if the application is chart dominant, which would save time and code for specific chart implementations, which is precisely how CAM functions. All test group developers found the CAM module helpful. Some expressed the value it brings to generating targeted description for specific types of chart. The senior developer on the test group expressed satisfaction with the loose coupling of CAM and MPAndroidChart, suggesting that it can easily be extended to other libraries. Most importantly, all feedback shows developer's lack of awareness of app accessibility and confirms the value of the proposed CAM tool.

## VI. Conclusion and Future Work

88.1% of charts are found to be inaccessible for vision-impaired users. Considering most developers have no vision-impaired issues, these issues go unnoticed. To bridge the gap of knowledge and reduce time costs, the CAM module is proposed which simplifies adding accessibility hooks to charts rendered using MPAndroidChart library. The approach is fully extensible to any third-party library. Developer testing showed that CAM reduced time costs and difficulty levels when implementing accessible charts, proving its effectiveness.

This aims to raise awareness in the developer community about charts accessibility. In future, this project can be extended in many ways. Individual exploration of chart components can be implemented to provide in-depth exploration of charts. Additional mediums like haptic and sonification of charts can further improve data perception. The hope is that this study opens up a new area of research which moves towards unblinding interactive charts.

# Appendix

**NOTE: All artifacts, code and result tables are *available on github* (*[https://github.com/ish-joshi/cam-chart-accessibility](https://github.com/ish-joshi/cam-chart-accessibility)*)**

## Developer Study Data

JD = Junior Developer SD = Senior Developer

| | Control | Test |
|---|---|---|
| Pie SD1 | No | Yes |
| Pie JD1 | No | Yes |
| Pie JD2 | No | Yes |
| | | |
| Bar SD1 | Yes | Yes |
| Bar JD1 | No | Yes |
| Bar JD2 | Yes | Yes |
| | | |
| Line SD1 | Yes | Yes |
| Line JD1 | Yes | Yes |
| Line JD2 | Yes | Yes |
| | | |
| Average Compl | 55.56% | 100.00% |
| | | |
| SD Completion | 66.67% | 100.00% |
| JD Completion | 50.00% | 100.00% |

| | Control | Test | | Average Time | Average control | Average Test |
|---|---|---|---|---|---|---|
| Pie SD1 | 20 | 12 | | | | |
| Pie JD1 | 20 | 14 | | | | |
| Pie JD2 | 20 | 7 | | 15.5 | 20.0 | 11.0 |
| | | | | | | |
| Bar SD1 | 15 | 9 | | | | |
| Bar JD1 | 20 | 13 | | | | |
| Bar JD2 | 18 | 9 | | 14.0 | 17.7 | 10.3 |
| | | | | | | |
| Line SD1 | 15 | 10 | | | | |
| Line JD1 | 20 | 11 | | | | |
| Line JD2 | 17 | 11 | | 14.0 | 17.3 | 10.7 |
| | | | | | | |
| Average time | 18.3 | 10.7 | | | | |
| | | | | | | |
| SD Average Tim | 16.7 | 10.3 | | | | |
| JS Average Tim | 19.2 | 10.8 | | | | |
| | | | | | | |
| | 2.5 | | | | | |

1 = not-accepate 5 = acceptable description
1 = hard 5 = easy
1 = not satisfied 5 = fully satisfied

| | Control | Test |
|---|---|---|
| Pie SD1 | 2 | 4 |
| Pie JD1 | 1 | 3 |
| Pie JD2 | 1 | 4 |
| | | |
| Bar SD1 | 3 | 4 |
| Bar JD1 | 2 | 4 |
| Bar JD2 | 2 | 5 |
| | | |
| Line SD1 | 2 | 4 |
| Line JD1 | 2 | 5 |
| Line JD2 | 2 | 5 |
| | | |
| Average Satisfa | 1.9 | 4.2 |
| | | |
| SD Average Tim | 2.3 | 4.0 |
| JS Average Tim | 1.7 | 4.3 |

| | Control | Test |
|---|---|---|
| Pie SD1 | 2 | 3 |
| Pie JD1 | 1 | 3 |
| Pie JD2 | 1 | 4 |
| | | |
| Bar SD1 | 3 | 4 |
| Bar JD1 | 1 | 3 |
| Bar JD2 | 2 | 4 |
| | | |
| Line SD1 | 2 | 4 |
| Line JD1 | 1 | 4 |
| Line JD2 | 2 | 5 |
| | | |
| Average difficul | 1.7 | 3.8 |
| | | |
| SD Average Tim | 2.3 | 3.7 |
| JS Average Tim | 1.3 | 3.8 |

| | Control | Test |
|---|---|---|
| Pie SD1 | 1 | 2 |
| Pie JD1 | 1 | 3 |
| Pie JD2 | 1 | 4 |
| | | |
| Bar SD1 | 1 | 3 |
| Bar JD1 | 1 | 4 |
| Bar JD2 | 2 | 5 |
| | | |
| Line SD1 | 3 | 3 |
| Line JD1 | 1 | 3 |
| Line JD2 | 1 | 4 |
| | | |
| Description Sco | 1.3 | 3.4 |
| | | |
| SD Average Tim | 1.7 | 2.7 |
| JS Average Tim | 1.2 | 3.8 |

# CAM Docs

```
# Purpose
CAM module is an independent module that can be plugged into existing data sources /
charting libraries to generate descriptions for common types of data.

## Structure
The `IDescriptor` interface is the common binding which has one `describe(): String` method
that must be implemented.

## Descriptors
Descriptor | Details
--- | ---
`BarChartDescriptor` | Use with bar charts that have categories on the X axis. For example,
car brand and sales is an idea example
`PieChartDescriptor` | Pie charts are described. Pass in proportions and labels and it will
generate a description. Alternatively title can be specified too.
`StockLineDescriptor` | Use this to describe stock charts with X axis being time and Y axis
being $ amount.
`RainfallColumnDescriptor` | Describes weather for each day of the week. Ideal for showing
weekly forecast.


## Usage
```kotlin
val barChartDescriptorDescription = BarChartDescriptor(
           arrayOf("Jazz", "City", "Accord", "HRV"),
           arrayOf(333f, 3223f, 234f, 342f),
           "Car model",
           "sale count",
           "Honda Car model sales count for 2020"
   ).describe()

   println(barChartDescriptorDescription)
```

## Implementing a custom descriptor
```kotlin
class ToImplementDescriptor : IDescriptor {

   // Do anything here...

   override fun describe(): String {
       return "generate a text description, which will be passed to screen-reader based on
state"
   }

}
```
```

## Usage with MPAndroidChart Modification

`getAccessibilityDescription()` is automatically called by the base `Chart.java` class when the screen-reader requests to populate an accessibility event when focused.

* Override the `getAccessibilityDescription()`
* Use a relevant descriptor or make your own
* Return the output of `describe()`

```java
class CustomPieChart extends PieChart {
/// other code goes here
   @Override
   public String getAccessibilityDescription() {

       // Gather all relevant data from chart for the descriptor
       PieData pieData = getData();
       IPieDataSet dataSet = pieData.getDataSetByIndex(0);
       String categoryTitleDataset = dataSet.getLabel();

       int entryCount = pieData.getEntryCount();

       Object[] labels = new Object[entryCount];
       Float[] proportions = new Float[entryCount];

       for (int i = 0; i < entryCount; i++) {
           PieEntry entry =dataSet.getEntryForIndex(i);
           float percentage = (entry.getValue() / pieData.getYValueSum());
           labels[i] = entry.getLabel();
           proportions[i] = percentage;
       }

       String categoryTitleMerged = categoryTitle.isEmpty() ? categoryTitleDataset :
categoryTitle;


       // Using chart state, generate the descriptor and describe it.
       IDescriptor pieChartDescriptor = new PieChartDescriptor(
               labels, proportions, categoryTitleMerged
       );


       return pieChartDescriptor.describe();
   }
}

```

```kotlin
```

```
private class CustomPieChart(context: Context) : PieChart(context) {

   override fun getAccessibilityDescription(): String {

       val pieData = data
       val dataSet = pieData.getDataSetByIndex(0)
       val categoryTitleDataset = dataSet.label

       val entryCount = pieData.entryCount

       val labels = arrayOfNulls<Any>(entryCount)
       val proportions = arrayOfNulls<Float>(entryCount)

       for (i in 0 until entryCount) {
           val entry = dataSet.getEntryForIndex(i)
           val percentage = entry.value / pieData.yValueSum
           labels[i] = entry.label
           proportions[i] = percentage
       }

       val pieDescriptor = PieDescriptor()


       return pieDescriptor.describe(labels, proportions, "category_title")
   }
}

// Optionally adding JVM overloads
class MyBarChart @JvmOverloads constructor(
       context: Context, attrs: AttributeSet? = null, defStyleAttr: Int = 0
) : BarChart(context, attrs, defStyleAttr) {

   // Code goes here.

}

```
```
```

# CAM Implementation

Only the *RainfallDescriptor is shown in detail.*

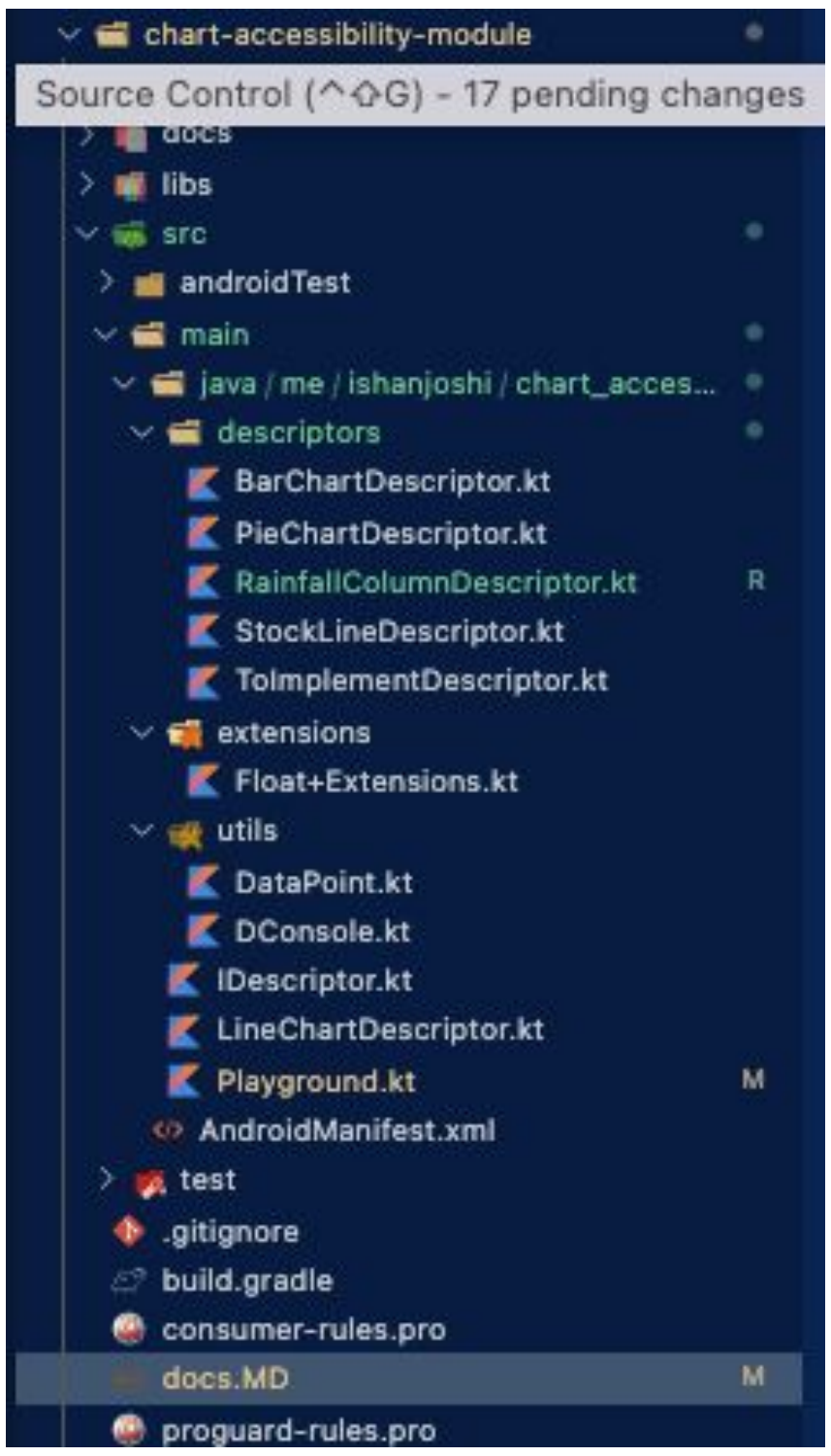

```
package me.ishanjoshi.chart_accessibility_module.descriptors

import me.ishanjoshi.chart_accessibility_module.IDescriptor
import me.ishanjoshi.chart_accessibility_module.utils.warn
import java.text.SimpleDateFormat
import java.util.*

data class RainfallDataPoint(
       val dateStamp: Date,
       val rainfall: Float
) {
   fun dayDescription(): String {
       val formatter = SimpleDateFormat("E dd MMM", Locale.getDefault())
       return formatter.format(dateStamp)
   }
```

```
   fun rainTextDescription(): String {
       val rfValue = "%.2f millimeters".format(rainfall)
       return when (rainfall) {
           0f -> "none"
           in 0f..2f -> "drizzle of $rfValue"
           in 2f..4f -> "light rain of $rfValue"
           in 5f..6f -> "moderate rain of $rfValue"
           in 8f..15f -> "moderate strong rain of $rfValue"
           in 15f..20f -> "strong rain of $rfValue"
           in 30f..700f -> "heavy rainfall of $rfValue"
           else -> rfValue
       }
   }
}

data class RainfallColumnDescriptor @JvmOverloads constructor(
       val epochTimeInWeek: Array<Long>,
       val rainfallMM: Array<Float>,
       val location: String? = null
) : IDescriptor {

   private val dataItems: List<RainfallDataPoint> = epochTimeInWeek.zip(rainfallMM).map {
       RainfallDataPoint(Date(it.first), it.second)
   }.filter { it.rainfall >= 0f } // only 0mm or more, cannot have negative rainfall.

   init {

       if (epochTimeInWeek.size != rainfallMM.size) {
           warn("Corresponding entries cannot be found, some entries may be omitted")
       } else if (dataItems.size != epochTimeInWeek.size) {
           warn("Cannot have negative rainfall")
       }

   }

   override fun describe(): String {

       val forLocation = if (location == null) "" else "for $location"

       val contextSetting = "This column chart describes the forecasted rainfall
$forLocation in the upcoming week. It has ${dataItems.size} entries"

       val eachDayDescription = dataItems.map {
           return@map "On ${it.dayDescription()}, ${it.rainTextDescription()} is
forecasted"
       }.joinToString(",")

       return "$contextSetting. $eachDayDescription"
   }
```

```
}
```

## Custom Data Tagging Script

```python
#  Copyright (c) 2020.
#
#  Ishan Joshi
#
#  This is part of Monash University Honours Project supervised by Dr
Chunyang Chen.
#

import os
import shutil
from infoextractor import InfoExtracter

APK_DIR = "apks"

# region ADB

from ppadb.client import Client as AdbClient

# Default is "127.0.0.1" and 5037
from ppadb.device import Device

ONEPLUS_SERIAL_NUMBER_ADB = "7d10f9f7"

client = AdbClient(host="127.0.0.1", port=5037)


def get_first_device() -> Device:
   print("Running lambda")
   devices = client.devices()
   return devices[0] if devices else None


# endregion

device = (lambda: get_first_device())()
assert device is not None, "No device found"


# assert device is not None, "Device is not available"


def get_relative_apk_path(apk_name):
   return f"{APK_DIR}/{apk_name}"


def check_valid_apk_name(apk_name):
   """
   :param apk_name: Checks if the apk_name exists in apks/{} directory
relataive
   :return: Boolean indication if it exists or not
   """
   return os.path.exists(get_relative_apk_path(apk_name))
```

```
def get_valid_chart_or_not() -> bool:
    ok = False
    inp = ""
    while not ok:
        inp = input("Has chart or no? y (has chart)/n (no
chart)?").lower()
        ok = inp in {"y", "n"}
        if not ok:
            print("Invalid.. try again")
    return inp == "y"


def get_type_of_chart():
    ok = False
    inp = ""
    valid = set("l,b,p".split(","))
    while not ok:
        inp = input("Type of chart? l (line), b (bar), p
(pie)?").lower()
        ok = inp in valid
        if not ok:
            print("Invalid.. try again")
    return inp


def get_is_accessible():
    ok = False
    inp = ""
    valid = set("y,n".split(","))
    while not ok:
        inp = input("Is it accessible? y (yes) / n (no)").lower()
        ok = inp in valid
        if not ok:
            print("Invalid.. try again")
    return inp


def loop_capture_data_until_end(apk_name, package_name):
    ended = False

    while not ended:
        # Ask input any key to log chart or not
        has_chart = get_valid_chart_or_not()
        extras = ""

        if has_chart:
            type_of_chart = get_type_of_chart()
            accessible = get_is_accessible()
            extras += f"y_{type_of_chart}_{accessible}"
        else:
            extras += "n_a_-"

        notes = input("Any additional notes?")

        device.uiautomator_snapshot(apk_name, package_name, extras,
notes)
```

```
        cont = input("Press any key to continue or type 'end' to stop
capture").lower()

        if cont == "end":
            ended = True

    print("Loop capture ended.... :)")


def enter_data_capture():
    # Get the apk file name
    apk_name = input("APK Name Please: ")
    if not check_valid_apk_name(apk_name):
        print(f"No apk found at path {get_relative_apk_path(apk_name)}.
Try again.")
        return

    if not apk_name.endswith(".apk"):
        print("Invalid file format. Must end with .apk")
        return

    print(f"Found APK file {apk_name}")

    print("Installing the app now")
    device.install(get_relative_apk_path(apk_name), reinstall=True,
grand_all_permissions=True)

    print("Installation complete")
    package_name =
InfoExtracter(get_relative_apk_path(apk_name)).package_name()

    device.launch_app(package_name)

    loop_capture_data_until_end(apk_name, package_name)

    input(f"type anything to uninstall package {package_name}")
    device.uninstall(package=package_name)

    print("Deleting file from apks folder...")
    os.remove(get_relative_apk_path(apk_name))


if __name__ == "__main__":
    ended = False

    enter_data_capture()
```

## PyADB Modifications

```
# Ishan Functions

    def launch_app(self, package_name: str):
        self.shell(f"monkey -p {package_name} -c android.intent.category.LAUNCHER 1")

    def get_xml_dump(self):
        out = self.shell(f"uiautomator dump /dev/tty")
        out = str(out).replace("UI hierchary dumped to: /dev/tty", "")
        return out

    def image_pull(self, saved_at, pull_to):
        process = subprocess.Popen(['adb', 'pull', saved_at, pull_to],
                                   stdout=subprocess.PIPE,
                                   stderr=subprocess.PIPE)
        stdout, stderr = process.communicate()
        stdout, stderr

    def __screenshot(self, file_name, out_path):
        """
        :param name: filename without the .png suffix
        :return: None
        """
        #  screencap -p /sdcard/screencap.png && adb pull /sdcard/screencap.png images/
        saved_at = f"/sdcard/{file_name}.png"
        self.shell(f"screencap -p {saved_at}")
        self.image_pull(saved_at, out_path)
```

```python
    def uiautomator_snapshot(self, directory_save, package_name,
additional_info, notes=""):

        package_name = package_name.replace(".", "-")

        out_dir_path = f"output/{directory_save}"
        if not os.path.exists(out_dir_path):
            os.makedirs(out_dir_path)

        fn =
f"{package_name}___{math.floor(time.time())}___{additional_info}"
        file_prefix = f"{out_dir_path}/{fn}"

        # Save the xml dump
        xml = self.get_xml_dump()
        with open(f"{file_prefix}.xml", 'w') as fh:
            fh.write(xml)
            fh.close()

        # Get screenshot
        self.__screenshot(fn, out_dir_path)

        if notes != "":
            with open(f"{file_prefix}.txt", 'w') as fh:
                fh.write(notes)
                fh.close()

        return "Done..."

    # Endregion
```

# MPAndroidChart Modifications

```
/**
     * Set this to true to enable "unbinding" of drawables.
When a View is detached
     * from a window. This helps avoid memory leaks.
     * Default: false
     * Link: http://stackoverflow.com/a/6779164/1590502
     *
     * @param enabled
     */
    public void setUnbindEnabled(boolean enabled) {
        this.mUnbind = enabled;
    }

    // region accessibility

    IDescriptor descriptor = new ToImplementDescriptor();

    public IDescriptor getDescriptor() {
        return descriptor;
    }

    public void setDescriptor(IDescriptor descriptor) {
        this.descriptor = descriptor;
    }

    /**
     * This method is called when the view is in focus, and it
is responsible for generating a textual description.
     * There may be a predefined behaviour that could suit the
requirements, but generally, overriding it is recommended.
     *
     * @return accessibility description must be created for
each chart
     */
    public String getAccessibilityDescription() {
        return descriptor.describe();
    }

    @Override
    public boolean
dispatchPopulateAccessibilityEvent(AccessibilityEvent event) {
        event.getText().add(getAccessibilityDescription());

        return true;
    }
```

PieChart.java — MPAndroidChart

PieChart.java

MPChartLib > src > main > java > com > github > mikephil > charting > charts > PieChart.java

```java

    @Override
    public String getAccessibilityDescription() {

        // Gather all relevant data from chart for the descriptor
        PieData pieData = getData();
        IPieDataSet dataSet = pieData.getDataSetByIndex(0);
        String categoryTitleDataset = dataSet.getLabel();

        int entryCount = pieData.getEntryCount();

        Object[] labels = new Object[entryCount];
        Float[] proportions = new Float[entryCount];

        for (int i = 0; i < entryCount; i++) {
            PieEntry entry =dataSet.getEntryForIndex(i);
            float percentage = (entry.getValue() / pieData.getYValueSum());
            labels[i] = entry.getLabel();
            proportions[i] = percentage;
        }

        String categoryTitleMerged = categoryTitle.isEmpty() ? categoryTitleDataset : categoryTitle;


        // Using chart state, generate the descriptor and describe it.
        IDescriptor pieChartDescriptor = new PieChartDescriptor(
                labels, proportions, categoryTitleMerged
        );


        return pieChartDescriptor.describe();
    }
}

```